\documentclass[review]{elsarticle}

\usepackage{lineno,hyperref}
\usepackage{multirow}
\usepackage{longtable}
\usepackage{graphicx}

\usepackage{algorithm}
\usepackage{algpseudocode}
\usepackage{float}
\usepackage{tikz}
\usepackage{amssymb}

\usepackage{natbib} 
\usepackage{mathtools}
\usepackage[all,cmtip]{xy}
\usepackage{caption}
\usepackage{subcaption}

\journal{EURO Journal on Computational Optimization}

\newtheorem{thm}{Theorem}

\newtheorem{proof}{Proof}
\newtheorem{fact}{Fact}

\long\def\comment#1\endcomment{}

\begin{document}
	
	\begin{frontmatter}
		
		\title{Reducing Dominating Sets in Graphs}
		
		\author{Ernesto Parra Inza\fnref{myfootnote2}}
		\ead{eparrainza@gmail.com}
		\fntext[myfootnote2]{ Centro de Investigación en Ciencias, UAEMor, Cuernavaca, Morelos, México.}
		
		\author{José María Sigarreta Almira\fnref{myfootnote3}}
		\ead{josemariasigarretaalmira@hotmail.com}
		\fntext[myfootnote3]{Facultad de Matemáticas, UAGro, Acapulco de Juárez, Guerrero, México.}
		
		\author[myfootnote2]{Nodari Vakhania\corref{mycorrespondingauthor}}
		\cortext[mycorrespondingauthor]{Corresponding author}
		\ead{nodari@uaem.mx}

		\begin{abstract}
A {\em dominating set} of a graph $G=(V,E)$ is a subset of vertices $S\subseteq V$
such that every vertex $v\in V\setminus S$ has at least one neighbor 
in set $S$. The corresponding optimization problem is known to be NP-hard. 
The best known polynomial time approximation algorithm for the problem separates
the solution process in two stages  applying first a fast greedy algorithm 
to obtain an initial dominating set, and then it uses an iterative procedure 
to reduce (purify) this dominating set. The purification stage 
turned out to be practically efficient. Here we further strengthen the 
purification stage presenting four new purification algorithms. All four 
purification procedures outperform the earlier purification procedure.
The algorithms were tested for over 1300 benchmark 
problem instances. Compared to the known upper bounds, the obtained solutions were
about 7 times better. Remarkably, for the 500 benchmark instances for which the optimum 
is known, the optimal solutions were obtained for 46.33\% of the tested instances, 
whereas the average error for the remaining instances was about 1.01.  		
		\end{abstract}
		
		
	\end{frontmatter}
	
	%\linenumbers
	
	%%%%%%%%%%%%%%%%%%%%%%%%%%%%%%%%%%%%%%%%%%%%%%%%%%%%%%%%%
	%%%%%%%%%%%%%%%%%%%%%%%%%%%%%%%%%%%%%%%%%%%%%%%%%%%%%%%%%
	
	\section{Introduction}

In a simple connected graph $G=(V,E)$, a subset of vertices $S \subseteq V$ is a \textit{dominating set} in graph $G$ if any vertex $v \in V$ is \textit{adjacent} to some vertex from this subset unless vertex $v$ itself belongs to set $S$.  The general objective is to find an \textit{optimal solution}, a dominating set with the minimum possible size $\gamma(G)$. The domination problem is known to be \textit{NP-hard} \cite{Garey} being among the hardest problems in the family.
It is closely related to other well-known graph optimization problems such as set cover, graph coloring and independent set. The problem of domination in graphs was formalized in  \cite{berge1962theory}  and \cite{ore1962theory}. Currently, this topic  has been detailed in the two well-known books by \cite{haynes1998domination} and \cite{haynes2017domination}. The  theory of domination in graphs is an area of  increasing interest in discrete mathematics and combinatorial computing,  in  particular, because of a number of important real-life applications. 
For example, in facility location problems in street networks  \cite{CORCORAN2021105368}, 
\cite{haynes2017domination}, \cite{JOSHI1994352} , in  monitor electric power systems with 
the aim of minimizing stage measurement units   \cite{liao2005power}, in 
electric power networks  \cite{haynes2002domination}, in the analysis of food web robustness
\cite{RTParra}. Different versions of graph domination
problem have been utilized to model clustering in wireless ad hoc networks  \cite{balasundaram2006graph}, \cite{wan2004distributed}, \cite{wu2003power}, and \cite{wu2002extended}. A number of other important graph-theoretic
problems including set cover, maximum independent set and chromatic number
are reducible  to the dominance problem  \cite{haynes2017domination}. Dominating sets
were also used in the study social networks  \cite{wang2009, wang2011} and \cite{abu2018}.

Due to the complexity status of the problem, the major efforts have been made towards the
development of heuristic algorithms. Among the earliest related works that we know 
we mention that of Chvatal \cite{chvatal}, who described an approximation algorithm 
with an approximation ratio $\ln\left(\frac{n}{\gamma}\right) + 1 + \frac{1}{\gamma}$ 
for the related 	weighted set cover problem; 
here and later $\gamma$ is the optimal objective value. 
Iteratively, the algorithm selects a vertex $v$  minimizing  
$\frac{w_v}{f(C\cup \{v\}) - f(C)}$, where $w_v$ is the weight of vertex $v$ and 
$f(C) = |\cup_{v\in C} N[v]|$, until $f(C) = |S|$. Adopting this algorithm, 	Parekh 
\cite{parekh} solved the  domination  problem showing that the cardinality of the 
dominating set created by his algorithm is upper bounded by $n+1-\sqrt{2m+1}$.  Later
Eubank et al. \cite{eubank} and Campan et al. \cite{campan} presented heuristic algorithms 
designed for special types of graphs estimating the performance of their algorithms
solely with with the experimental study. Recently, \cite{tcs22} described a 
heuristic algorithm with a better performance. 

A little research has been done in the development of exact algorithms. To the best of our
knowledge, the only exact algorithms which do better than a complete enumeration were 
described in \cite{van2011exact}, \cite{iwata2012faster} and \cite{Parra2022}. In \cite{van2011exact} and \cite{iwata2012faster}, the authors show that their
algorithms run in times $O(1,4969^{n})$ and $O(1.4689^{n})$, respectively, without 
presenting an experimental evidence for the practical performance of their algorithms. 
Although these bounds are notably better than $2^{n}$, they remain too impractical.
Quite recently, in   \cite{Parra2022} two exact algorithms, an implicit enumeration 
algorithm and an alternative integer linear programming (ILP) formulation 
were proposed. The authors also
describe an approximation algorithm. On the one hand, it gave the solutions
of notably better quality than the previously known approximation algorithms 
for the benchmark instances with up to 2100 vertices, some of them being solved
in less than 1 minute. The algorithm found an optimal solution for 61.54\% of these 
instances. On the other hand, it failed to create solutions within a 
reasonable time limit for the graphs with 3000 and more vertices. In opposit to
this, for large-scale practical problems, the heuristic from \cite{tcs22} gave 
solutions within the time limit of 50 seconds for the graphs with up to 14000 vertices. 

In this paper,  relaying on the basic framework of the above heuristic,  
we describe new procedures that further improve the quality of the solutions for 
large-scaled instances. The heuristic works in the two stages. Similarly to the 
heuristic from  \cite{tcs22}, in stage 1, a dominant set is generated, which is
reduced or {\em purified} at stage 2. This purification stage turned out to be
particularly effective, in practice. The reduction is achieved through the analysis of the 
flowchart of stage 1 combined with a special kind of clustering of the dominating set 
of stage 1. A cluster exploits a especial structural relationship between the vertices
of a dominating set, that can beneficially be used during its purification process. 
Important dependencies in the dominant set represented by a cluster permit us
to purify it in different ways. In particular, the order, in which the vertices of 
each cluster are purified is important and affects the outcome of the purification process.
A cluster can be seen as a specific type of spanning forest of the initial graph $G$. 
Different clustering methods  yield a collection of different types of rooted trees.
By a traversal of these trees, the initial dominating set is purified at stage 2. 
Different combinations of a particular clustering and traversal/purification method 
result in different overall purification algorithms of different efficiency.

Exploring further the structural properties of a dominating set, we propose a new 
clustering method for a more beneficial analysis of the structure of the dominating set. 
We describe different traversal methods for the new cluster structure leading to
four different purification procedures, all of them outperforming the purification 
procedure from \cite{tcs22}. For over 1300 benchmark problem instances  \cite{bdparra2}, 
the improvement was about 10\% on average. Compared to the 
known upper bounds, the obtained solutions, on average, are about 7 times better
resulting in the reduction of 85.71\% of the value of the best of these upper bounds. 
For the 500 benchmark instances from \cite{bdparra2} where the optimum is known 
(see also \cite{Parra2022}), at least one of the new purification  procedures 
obtained an optimal solution for 46.33\% of the instances, whereas an average 
error for the remained instances was 1.01.	

The outline of the remaining part of the paper is as follows. 
In the following Section 2, we describe the necessary preliminaries. In Section 3, we 
give our clustering procedure. Section 4 details the four purification procedures. In Section 5, 
we present the approximation ratios, and in Section 6 we report our experimental results 
in detail and give some concluding remarks.

%%%%%%%%%%%%%%%%%%%%%%%%%%%%%%%%%%%%%%%%%%%%%%%%%%%%	
%%%%%%%%%%%%%%%%%%%%%%%%%%%%%%%%%%%%%%%%%%%%%%%%%%%%

\section{Preliminaries}

A formal description of our domination problem is as follow. Given 
a simple connected graph $G=(V,E)$ with $|V|=n$ vertices and $|E|=m$ edges, a subset of vertices $S \subseteq V$ is a \textit{dominating set} in graph $G$ if any vertex $v \in V$ is \textit{adjacent} to some vertex $x$ from this subset (i.e., there is an edge $(v,x)\in E)$ unless vertex $v$ itself belongs to set $S$. The objective is to find a feasible solution with the minimum possible size $\gamma(G)$. 

Given a vertex $v\in V$, $N(v)$ is the set of neighbors or the open neighborhood of $v$ in $G$; that is,  $N(v)=\{u\in V: (u,v)\in E\}$. We denote by $\delta_G(v)=|N(v)|$  the \emph{degree} of vertex $v$ in $G$. We let $\delta_G = \min_{v\in V}\{\delta(v)\}$ and $\Delta_G = \max_{v\in V}\{\delta(v)\}$. The \emph{private neighborhood} of vertex $v \in S\subseteq V$ is defined by $\{u \in V: N(u)\cap S = \{v\}\}$; a vertex in the private neighborhood of vertex $v$ is said to be its \emph{private neighbor} with respect to set $S$. For further details on the basic graph terminology,
 see \cite{haynes2017domination}.

In \cite{tcs22} a dominating set is formed in two stages. At the first stage an 
initial dominating set $S = \{v_1,\ldots, v_k\}$ is created by a fast 
greedy algorithm, where $v_i$ stands for the vertex 
included at iteration $i$ of this greedy.  At the second purification stage, 
a special iterative procedure is applied to reduce the initial dominating 
set. This iterative procedure will be referred to as the 
purification procedure 0, abbreviated $PP_0$. It 
is based on the analysis of the flowchart of the greedy, 
represented as a special kind of a spanning forest $T$ of the original 
graph $G$  and consisting of the vertices of set $S$. This forest is the union 
of the so-called clusters which are formed at the first stage while greedy 
generates set $S$. 

The purification process is accomplished during a traversal of each cluster.
Combination of different clustering and traversal methods yield different
purification procedures.

\section{Stage 1: The clustering}

We organize vertices from set $S$ in special graphical structures, the
so-called {\em clusters}. A cluster $C_i$ is a rooted tree, a sub-graph of 
graph $G$ associated with some connected component of the graph induced by 
set $S$. Initially, a set of clusters ${C_1,\dots, C_l}$ form a partition of set 
$S$. We denote the union of these trees by $T$. 
In a cluster, the vertices of each connected component can be represented 
in different ways as trees. During the traversal of these trees, special 
conditions are verified and some vertices are omitted (purified). 

We generate the clusters during the construction of the initial dominating set 
$S$. This construction is based on the greedy procedure from  \cite{tcs22} 
further referred to as {\em Greedy}. It works in a number of iterations, which
number is upper bounded by $n$. At each iteration 
$h\geq 1$, one specially selected vertex, denoted by $v_h$, is added to the dominant
set $S_{h-1}$ of the previous iteration, i.e.,  $S_h:=S_{h-1}\cup\{v_h\}$, 
where initially $S_0=\emptyset$. $S^{h-1}$ is the partial dominant set of iteration $h-1$.
At each iteration $ h $, the \textit{active degree} of a
vertex $ v \in \overline{S_{h-1}}  $ is 
 $ | N(v) \setminus [S_{h-1} \cup N(S_{h-1})]| $. At each iteration $ h $, vertex
$ v_h $  is a vertex of $ \overline{S_{h-1}} $ with the maximum active degree.
Note that $v_1$ is a vertex with the maximum degree in graph $G$. 
Greedy halts when $ S_h $ is a dominating set of $ G $. At that iteration, all 
uncovered vertices with active degree 0 (if any) are included in the set $S_h$.
A formal description is given below. 

\medskip

\begin{algorithm*}
	\caption{Algorithm Greedy}\label{alg_basic}
	\begin{algorithmic}
		\State Input: A graph $G$.
		\State Output: A dominating set $ S $ of $ G $.
		\State $h := 1$; 
		\State $i := 1$;
		
		\State $S_0:= \emptyset$;
		\State $T^0:=\emptyset$;  
		
		\{ iterative step \} 
		
		\While{$S_h$ is not a dominating set of  graph $G$} 
		\State $h := h+1$; 
		\State $v_h$ := any vertex of $ \overline{S_{h-1}} $ with maximum active degree;  
		\State $S_h := S_{h-1}\cup \{v_h\}$;
		\State $[T^h,i] := $Cluster\_Generation($v_h$, $ T^{h-1}$, $h$, $i$);
		\EndWhile
		
	\end{algorithmic}
	
\end{algorithm*}

\medskip

\newpage

Throughout this section, let $v_h$ is used for the vertex with the maximum active degree
selected by {\em Greedy} at iteration $h$, hence  $S = \{v_1,\ldots, v_k\}$ is the dominant set of Stage 1.

\begin{figure}[H]
	\centering
	\includegraphics[width=1.0\linewidth]{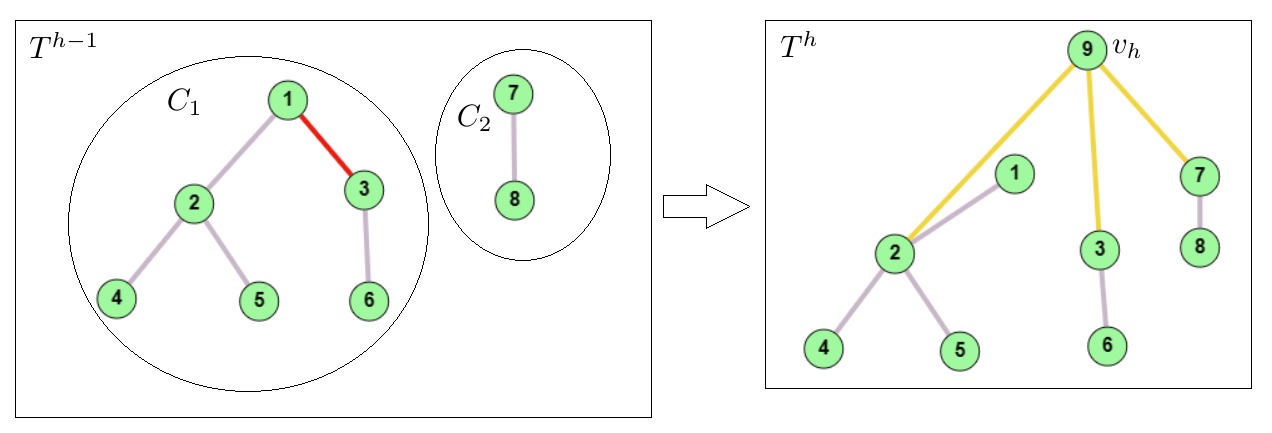}
	\caption{Forests $T^{h-1}$ and $T^{h}$ 
	(the red edge in $T^{h-1}$ is omitted in $T^{h}$}
	\label{figura1}
\end{figure}

Our aim is {\em purify} 
this set, i.e., omit some vertices from that set so that the reduced set 
remains dominant. A combination of a particular clustering and traversal 
methods results in a particular purification procedure. 
	The four purification procedures that we propose here apply the same rules for
	the creation of the clusters. 
	Unlike the purification procedure $PP_0$, 
	which was oriented
	on the creation of clusters in depth, here Stage 1 creates clusters in
	width.  Iteratively, every next vertex $v_h$ is included in the partial 
	forest $T^{h-1}$ at iteration $h$, $h=1,\dots,k$, as follows:  
	
	\begin{itemize}
	 
	 \item Initially, we let $T^0:=\emptyset$ and $T^1:=C_1:=v_1$
	    
	 \item For $h>1$,  	
  \begin{itemize}
		\item If $N_{V(T^{h-1})}(v_h)=\emptyset$, then we create a new 
		cluster $C_i:=v_h$  
		and \{update $T^h$\} $T^h:=T^{h-1} \cup C_i$ 		
		
		\item If $N_{V(T^{h-1})}(v_h)=\{x\}$, $x \in T$ (i.e., $|N_{V(T^{h-1})}(v_h)|=1$), 
		then $T^h$ is $T^{h-1}$ complemented by vertex $v_h$, a new child of vertex $x$ 
		
		\item If $|N_{V(T^{h-1})}(v_h)|>1$, we create a new cluster $C_i$ with root  $v_h$ and with the immediate successors from $N_{V(T^{h-1})}(v_h)$. In this case cluster $C_i$ includes one or more clusters from $T^{h-1}$ (these clusters disappear as individual clusters in forest  $T^h$). This kind of a union may yield a cycle
		(see Figure  \ref{figura1}). To avoid this, we eliminate a required
		number of edges from $T^{h-1}$ incident with vertices in $N_{V(T^{h-1})}(v_h)$. 
			\end{itemize}
\end{itemize}

 Below we give a formal description of the procedure.

\medskip
	
	\begin{algorithm*}[h!]
		\caption{Cluster\_Generation}\label{cluster}
		%\fontsize{7pt}{.2cm}\selectfont
		\begin{algorithmic}
			\State Input: $v_h$, $ T^{h-1}$, $h$, and number of clusters created $i$.
			\State Output:  Forest $T^h$ and $i$ .
			%\State
			\If{$h=1$}			
			\State $C_1 := \{v_1\}$; \{$v_1$ is root of $C_1$\}
			\State $T^1 := C_1$;
			\Else
			%\State $i := 2$;
			
			%\For {any $v_h \in S$, $h>1$} \hspace{.3cm} \{in order\}
						
			\If {$N_{V(T^{h-1})}(v_h) = \emptyset$}
			\State $i := i + 1$;
			\State $C_i := \{v_h\}$; \{$v_h$ is root of $C_i$\}			
			\State $T^{h} := T^{h-1} \cup C_i$;						
			
			\Else  
			
			\If{$|N_{V(T^{h-1})}(v_h)| = 1$}
			\State $x = N_{V(T^{h-1})}(v_h)$
			\State $T^{h} := T^{h-1} \cup \{x\}$   \{insert $ v_h $ as a child of $ x $\};
			\Else
			\State $T^{h} := T^{h-1} \cup N_{V(T^{h-1})}(v_h)$    \{insert $ N_{V(T^{h-1})}(v_h) $ as a child of $ v_h $ and update $T^{h}$ \};
			%\State Update $T^{h}$;
			\EndIf
			
			\EndIf
			\EndIf
		\end{algorithmic}
		
	\end{algorithm*}
\medskip

\newpage 

\medskip
		
	\begin{thm}\label{time-c}
	Stage 1 runs in time $ O(n^3) $. 
	\end{thm}
	\begin{proof}
	The number of trees $C_i$ and the number of vertices in each tree 
		are clearly bounded 	by $|V(T)| < n$. At an iteration $h$,  $N_{V(T)}(v_h)$ can be determined in time $O(|V(T)|)$, and the same time is required to locate and update each vertex 	$x \in N_{V(T)}(v_h)$. Hence, at every iteration $h$,  vertex $v_h$ can be
	incorporated into the forest $T^{h-1}$ in time $O(|V(T)|*|N_{V(T)}(v_h)|)=O(n^2)$.
	Theorem follows as in Greedy, vertex $v_h$ is determined and added in the dominating set $S$ also in time $O(n^2)$ (see \cite{tcs22}).  
	\end{proof}

	%%%%%%%%%%%%%%%%%%%%%%%%%%%%%%%%%%%%%%%%%%%%%%%%%%%%%
	%%%%%%%%%%%%%%%%%%%%%%%%%%%%%%%%%%%%%%%%%%%%%%%%%%%%%

	\section{Stage 2: Purification procedures} 
		
Given a set of clusters formed during the execution of the algorithm of Stage 1, 
   we wish to purify the forest $T$. 		
	The purification procedure $PP_0$,  
	at Stage 1, during the 
	construction of the initial dominating set $S = \{v_1,\ldots, v_k\}$, 
	inserts vertex  $v_h$ in the current forest $T^{h-1}$ as a child of 
	one of the earlier inserted vertices $v\in T^{h-1}$ ($(v,v_h) \in E(G)$).
	Vertex $v\in C$  ($C\in T^{h-1}$) is one which is furthest away from the 
	root in its cluster. In case the root of another cluster from $T^{h-1}$ is
	adjacent to vertex $v_h$ (in graph $G$), this root will become a son
	of $v_h$. Then note that the clusters with such roots will be merged
	with the former cluster $C$ into a new larger cluster.

Thus in the procedure $PP_0$,  
the clusters are generated 
in depth resulting in clusters with high depth. Dealing with 
high clusters has some advantages but also some disadvantages. 
In the purification procedure $PP_0$, 
at Stage 2, each
cluster is traversed in a bottom up fashion, from a leave to the root, and a quadruple $(a,b,c,d)$ or a trio $(a,b,c)$, forming sub-parts of the  corresponding path, are identified. Given a quadruple
$(a,b,c,d)$, vertices $(b,c)$ are purified unless such vertex possesses a 
{\it semi-private neighbor} (a vertex 	$x\in V (G) \setminus T^h$ is a 
semi-private neighbor of vertex $v\in T^h$ if it
is the only (remained) neighbor of vertex $x$ from $T^h$). For a trio, 
similar condition is verified  only for vertex $b$. 
	
As above described, two vertices can be purified at once. However, this might not be possible, either because there may exist no quadruple (the up going paths might be too short)	or/and because a candidate vertex may possess a semi-private neighbor. It might be beneficial to set firm all vertices with a semi-private neighbor first 
(a firm vertex remains unpurified in the resultant dominant set). It
may also be beneficial to leave a vertex possessing a considerable number of 
immediate descendants without semi-private neighbors unpurified in case all 
its immediate descendants are purified. Likewise, it might be 
good to leave a vertex unpurified  if the number of its neighbors in   
$ \overline{S} $ is ``large enough''. The above and similar considerations were
taken into account in four new purification procedures.

%%%%%%%%%%%%%%%%%%%%%%%%%%%%%%%%%%%

The first three  procedures 1-3 set firm all vertices of forest $T$ with private neighbors at iteration 0. If the so formed set is dominating, all  stop.  
Otherwise, three different disjoint subsets of set $S$ forming a partition of $S$
are distinguished, as defined below.
 
\begin{enumerate}
\item[(1)] The set of firm vertices $F^h$ (as noted above, $F^0$ is 
the set of vertices from $T$ with private neighbors, 
at iteration $h>0$, $F^h$ is the set of vertices from $T$ with \textit{semi-private} neighbors). 

\item[(2)] The set of the pending vertices $P^h$ formed by yet non-firm vertices which
are neighbors of a firm vertex.
 
\item[(3)] Set $U^h$ consisting of the remaining yet unconsidered vertices. 
\end{enumerate}

Iteratively, once $F^h$ is determined, sets $P^h$ and $U^h$ are formed. 
The two purification parameters for each cluster vertex are defined below. 

The {\it outer cover set} of vertex $v\in T^{h-1}$ at iteration $h$, OCS$(v,h)$, is the set of  neighbors of $v$ from $V\setminus S$, which are not adjacent with any
firm vertex from set $F^{h-1}$; i.e., $OCS(v,h)=N_{\overline{T^{h-1}}}(v)\setminus N_{\overline{S}}(F^{h-1})$. 

The {\it inner cover set} of vertex $v\in T^{h-1}$ at iteration $h$, ICS$(v,h)$, 
is the set of yet non-firm neighbors of vertex $v$ in $T^{h-1}$, i.e., 
$ICS(v,h)=N_{T^{h-1}}(v)\setminus F^{h-1}. $

Roughly, vertices with high $|OCS(v,h)|$ and $|ICS(v,h)|$  tend to be non-purified;  visa-versa, direct descendants of such a vertex with a low outer and inner cover indices tend to be purified. An overall purification balance of vertex $v$  is % cambiar formula! 
 $PB(v,h)=\alpha |OCS(v,h)| + \beta |ICS(v,h)|$, where $0\le \alpha,\beta \le 1$.

%%%%%%%%%%%%%%%%%%%%%%%%%%%%%%%%%%%%%%%%%%%%%%%%%%%%%%%%%%%%%%%%%%%%%%%%%%%%%%%%%%%%%%%%%%%%%%%%%%%%%%%%%%%%%%%%%%%%%%%%%
The distinguished features of our  procedures are defined below. 
	
{\bf Procedure 1.} 
At iteration $h>0$, for every cluster $C\in T$ is traversed in a bottom-up 
fashion. For every non-leaf vertex $x$ at a given level of cluster $C$,  
if $x$ is firm, its children without semi-private neighbors are purified. 
If $x$ is not firm and has no firm children, it is set firm and its children 
with no semi-private neighbors are purified, whereas the children with semi-private neighbors are set firm. If $x$ has a firm child, the parent of $x$ is set firm
(unless it is already firm) and $x$ is purified:  
	
	\begin{algorithm*}[]
		\caption{Procedure Purify 1 ($PP_1$). }\label{alg_purification1_stage}
		\fontsize{10pt}{.35cm}\selectfont
		\begin{algorithmic}
			\State Input: Forest $T= \{C_1, C_2, ..., C_k\}$.
			\State Output: $S^*$. \hspace{.3cm} \{ minimal dominating set \}
			
			\{set vertices with private neighbors\}	
			
			\State $F^0 :=$ \{set of vertices from $T$ with private neighbors\}; % $ \emptyset$;

			\If {$F^0$ is dominating set} 
			\State $S^*:=Firm$;
			\Else
			\For {any $C_i \in T$} \hspace{.3cm} \{ in order \}
			\State $C^0_i := C_i$;
			\State $h := 0$;
			\State\{Travel $ C_i $ by levels in ascending order\}
			\For {any $x\notin$ leaf set in level $j$ of $C_i$}
			\If {$x \in F^h$} \hspace{.3cm} 
			\State $C^{h+1}_i := C^h_i\setminus \{$ children not a firm of $x$ do not have semi-private neighbors$\}$; %\hspace{.3cm} \{ Purify non firm children 
			\Else
			\If{$x$ has non firm children}
			\State $F^{h+1} := F^{h}\cup \{x\}$;
			\State $C^{h+1}_i := C^h_i\setminus \{$children of $x$ do not have semi-private neighbors$\}$; %\hspace{.3cm} \{ Purify children of $x$\}
			\Else 
			\If{parent of $x$ is not in $F^{h}$}
			\State $F^{h+1} := F^{h}\cup \{$ parent of $x\}$;
			\State $C^{h+1}_i := C^h_i\setminus \{x\}$;
			\Else
			\State $C^{h+1}_i := C^h_i\setminus \{x\}$;
			\EndIf
			\EndIf
			\EndIf
			\State $h := h+1$; 
			\EndFor
			
			\EndFor
			\State $S^*:=F^{h}$;
			\EndIf
			
		\end{algorithmic}
	\end{algorithm*} 
	\newpage
	
%%%%%%%%%%%%%%%%%%%%%%%%%%%%%%%%%%%%%%%%%%%%%%%%%%%%%%%
{\bf Procedure 2.}
At an iteration $h>1$, non-firm vertex $x$ with the maximum  $PB(x,h)$ is set firm
(for each non-firm vertex $v$, the sets OCS$(v,h)$ and ICS$(v,h)$ being updated respectively). The procedure halts at iteration $h$ if $U^h=\emptyset$  and outputs
dominant set $F^h$:

	\begin{algorithm*}[]
		\caption{Procedure Purify 2 ($PP_2$). }\label{alg_purification2_stage}
		\fontsize{10pt}{.35cm}\selectfont
		\begin{algorithmic}
			\State Input: Forest $T= \{C_1, C_2, ..., C_k\}$.
			\State Output: $S^*$. \hspace{.3cm} \{ minimal dominating set \}
			
			%\State $F^0 := \emptyset$;  
			\State $F^0 :=$ \{set of vertices from $T$ with private neighbors\}; % $ \emptyset$;
			\State $P^0 := \emptyset$;

			\If {$F^0$ is dominating set} 
			\State $S^*:=F^0$;
			\Else
			\State $h:=0$;
			\For {any $C_i \in T$} \hspace{.3cm} \{ in order \}
			\While{$V(C_i)\setminus (F^h\cup P^h)\neq\emptyset$} 
			\State $v$ := any vertex with the maximum PB$(v,h)$ in set $V(C_i)\setminus F^h$; 
			\State $F^{h+1} := F^h\cup \{v\}$;
			\State $P^{h+1} := P^h\cup (N_{V(C_i)}(v)\setminus F^h)$;
			\State $h:=h+1$;
			\EndWhile
			\If {$F^h$ is dominating set} 
			\State $S^*:=F^h$;
			\State break;
			\EndIf
			
			\EndFor
			\EndIf
			
		\end{algorithmic}
		
	\end{algorithm*} 
	
	%%%%%%%%%%%%%%%%%%%%%%%%%%%%%%%%%%%%%%%%%%%%%%%%%%%%%%%%%%%%%%%%%%%%%%%%%%%%%
	
{\bf Procedure 3.}
At an iteration $h>1$, a vertex $x$ with the minimum $PB(x,h)$ is  purified and the
vertices from $T^{h-1}$ with semi-private neighbors from $\overline T^{h-1}$ 
are set firm (for each non-firm vertex $v$, the sets OCS$(v,h)$ and ICS$(v,h)$ being updated respectively). The procedure halts at iteration $h$ if $U^h=\emptyset$  and outputs dominant set $F^h$:

	\begin{algorithm*}[h!]
		\caption{Procedure Purify 3 ($PP_3$). }\label{alg_purification3_stage}
		\fontsize{10pt}{.35cm}\selectfont
		\begin{algorithmic}
			\State Input: Forest $T= \{T(C_1), T(C_2), ..., T(C_k)\}$.
			\State Output: $S^*$. \hspace{.3cm} \{ minimal dominating set \}

			\State $F^0 :=$ \{set of vertices from $T$ with private neighbors\}; % $ \emptyset$;
			\State $P^0 := \emptyset$;

			\If {$F^0$ is dominating set} 
			\State $S^*:=F^0$;
			\Else
			\State $h:=0$;
			\For {any $T(C_i) \in T$} \hspace{.3cm} \{ in order \}
			\While{$T(C_i)\setminus (F^h\cup P^h)\neq\emptyset$} 
			\State $v$ := any vertex with the minimum PB$(v,h)$ in set $T(C_i)\setminus F^h$; 
			
			\State $P^{h+1} := P^h\cup \{v\}$;
			\State $F^{h+1} := F^h\cup \{$all vetex in set $T\setminus (F^h\cup P^{h+1})$ that have semi-private neighbor $\}$;

			\State $h:=h+1$;
			\EndWhile
			\If {$F^h$ is dominating set} 
			\State $S^*:=F^h$;
			\State break;
			\EndIf
			
			\EndFor
			\EndIf
			
		\end{algorithmic}
		
	\end{algorithm*} 
	\newpage
	
	%%%%%%%%%%%%%%%%%%%%%%%%%%%%%%%%%%%%%%%%%%%%%%%%%%%%%%%%%%%%%%%%%%%%%%%
{\bf Procedure 4.}	
One of good intuitions in the purification procedure ($PP_0$) %propoced in \cite{tcs22} 
is the following: it might be convenient that, for any adjacent nodes $a$, $b$ of $T_i$ such that node $a$ is a leaf of $T_i$, node $ a $ is processed earlier than node $ b $. Then, while going through such a good intuituion, it seems that the kernel is the following: it might be convenient that, for any adjacent vertices $ a $, $ b $ of $ S $ such that $ a $ was added to $ S $ [in Stage 1 \cite{tcs22}] later than $ b $, vertex $ a $ is processed earlier than vertex $ b $  [in fact vertex $ a $ might have a \textit{private neighbor} in $V \setminus S$ with respect to vertex $ b $]. That leads to the following possible alternative procedure for purification.
	
\textit{Procedure 4 ($PP_4$)}: Let us write $S=\{v_1, ..., v_p\}$ where $(v_1, ..., v_p)$ is the total order given by the reverse ordering in which such vertices were added to $ S $ [in Stage 1 \cite{tcs22}]; set $S^{*} := S$; for $i = 1, ..., p$, if $N[v_i] \subseteq (S^* \setminus \{v_i\})\cup N(S^* \setminus \{v_i\})$, then set $S^{*} := S^* \setminus \{v_i\}$ [i.e. purify/remove $v_i$ from $S$]; return $S^{*}$.

This procedure was  basically suggested by one of the anonymous referees of 
 \cite{tcs22}. The authors highly acknowledge  this suggestion.
	
	%%%%%%%%%%%%%%%%%%%%%%%%%%%%%%%%%%%%%%%%%%%%%%%%%%%%%%%%%%%%%%%%%
\section{Approximation ratios}
	In this section we derive approximation ratios for our algorithms. By \cite{tcs22}, if $|S| \leq 2$ then $S$ is a minimum dominating set. Then, all the graphs $G$ analyzed here will be such that  $\gamma(G) > 2$. Note that $G$ is a connected graph, and all $x\in S^*$ has at least one semi-private neighbor ($S^{*}$ be the output of purification procedures). Then $|S^{*}| \leq n/2$.
	
	In the next, we derive an approximation ratio of purification procedures. By \cite{haynes2017domination}, $\gamma(G) \geq \frac{n}{\Delta+1}$.  
	Then $\frac{n}{\Delta+1} \leq |S^*| \leq \frac{n}{2}$. Since $|S^*|\geq \gamma(G)$, we obtain that an approximation ratio for purification is $\displaystyle\rho = \frac{|S^*|}{\gamma(G)} \leq \frac{\frac{n}{2}}{\frac{n}{\Delta+1}} = \frac{\Delta+1}{2}$.
		
	By \cite{tcs22}, Algorithm Greedy  achieves the approximation ratio $\rho \leq  ln(\Delta+1)+1$. Using these approximation ratios, we obtain an overall approximation ratio for purification procedures.
	
	\begin{center}
		$\rho \leq \left\{\begin{array}{ll}
			\frac{\Delta + 1}{2}, & \mbox{if 1 $\leq \Delta \leq$ 4} \\[.3cm]
			ln(\Delta+1)+1, & \mbox{otherwise.}
		\end{array}
		\right.$
	\end{center}

	%%%%%%%%%%%%%%%%%%%%%%%%%%%%%%%%%%%%%%%%%%%%%%%%%%%%%%%%%%%%%%%%%%%%%%%%%
	%%%%%%%%%%%%%%%%%%%%%%%%%%%%%%%%%%%%%%%%%%%%%%%%%%%%%%%%%%%%%%%%%%%%%%%%%
	
	\section{Experimental results}
	
	In this section we describe our computation experiments. We implemented our algorithms in C++ using Windows 10 operative system for 64 bits on a personal computer with Intel Core i7-9750H (2.6 GHz) and 16 GB of RAM DDR4. We generated different 
	sets of problem instances using different pseudo-random methods for generation
	of graphs.

	We analyzed  1316 benchmark instances from \cite{bdparra2}. A complete 
	summary of the results for these instances can be found at \cite{bdparra2}.
	Due to the space limitations, here we summarize our results for 45 sample       
	randomly selected instances in Tables \ref{table1} and \ref{table3}, where $S$ denotes
	the initial dominant set found by \cite{tcs22}, and $S^*$ is the purified
	dominant set returned by the corresponding purification procedures. The
	columns marked by \% represent the percentage of the reduction of the 
	number of vertices from the dominant set returned by the $PP_0$ % Purification Procedure 0 
	in the dominant set returned by the purification procedures 1-4. 
	
   Table \ref{table1} below presents the percentage of the improvement by
   purification procedures 1-4 for the 45 sample instances with sizes between
	800 and 15000. 

	%\newpage
	
	% Please add the following required packages to your document preamble:
	% \usepackage{multirow}
	% \usepackage{longtable}
	% Note: It may be necessary to compile the document several times to get a multi-page table to line up properly
	%\footnotesize%%%%%%%%%%%  smaller font size %%%%%%%%
	\scriptsize
	\begin{longtable}[c]{cccccccccccc}
%		\caption{Results of purification procedures and percent improvement}
%		\label{table1}\\
		\hline
		\multirow{3}{*}{\textbf{$|V|$}} & \multirow{3}{*}{\textbf{$|E|$}} & \multicolumn{2}{c}{\multirow{2}{*}{$PP_0$}} & \multicolumn{8}{c}{\textbf{Purification Procedures}} \\ \cline{5-12} 
		  &  & \multicolumn{2}{c}{} & \multicolumn{2}{c}{$PP_1$} & \multicolumn{2}{c}{$PP_2$} & \multicolumn{2}{c}{$PP_3$} & \multicolumn{2}{c}{$PP_4$} \\ \cline{3-12} 
		  &  & \textbf{$|S|$} & \textbf{$|S^*|$} & \textbf{$|S^*|$} & \textbf{\%} & \textbf{$|S^*|$} & \textbf{\%} & \textbf{$|S^*|$} & \textbf{\%} & \textbf{$|S^*|$} & \textbf{\%} \\ \hline
		\endhead
		\hline
		\endfoot
		\endlastfoot
		 840 & 310396 & 7 & 5 & 4 & 20.00 & 4 & 20.00 & 5 & 0.00 & 4 & 20.00 \\
		 1150 & 1185 & 500 & 470 & 431 & 8.30 & 430 & 8.51 & 430 & 8.51 & 428 & 8.94 \\
		 1650 & 1711 & 717 & 660 & 614 & 6.97 & 609 & 7.73 & 611 & 7.42 & 607 & 8.03 \\
		 1700 & 1740 & 725 & 681 & 634 & 6.90 & 632 & 7.20 & 633 & 7.05 & 632 & 7.20 \\
		%5 & 1800 & 1905 & 751 & 715 & 663 & 7.27 & 660 & 7.69 & 663 & 7.27 & 660 & 7.69 \\
		 1900 & 1977 & 802 & 754 & 702 & 6.90 & 698 & 7.43 & 703 & 6.76 & 697 & 7.56 \\
%		7 & 2050 & 2199 & 868 & 810 & 755 & 6.79 & 771 & 4.81 & 752 & 7.16 & 747 & 7.78 \\
		 2150 & 2287 & 909 & 850 & 787 & 7.41 & 781 & 8.12 & 784 & 7.76 & 781 & 8.12 \\
		 2300 & 2407 & 966 & 907 & 842 & 7.17 & 835 & 7.94 & 841 & 7.28 & 835 & 7.94 \\
		%10 & 2350 & 2391 & 1042 & 966 & 898 & 7.04 & 890 & 7.87 & 892 & 7.66 & 890 & 7.87 \\
		 2450 & 2588 & 1054 & 987 & 901 & 8.71 & 895 & 9.32 & 898 & 9.02 & 895 & 9.32 \\
%		12 & 2550 & 2583 & 1099 & 1022 & 950 & 7.05 & 939 & 8.12 & 944 & 7.63 & 939 & 8.12 \\
%		 2600 & 2681 & 1131 & 1051 & 979 & 6.85 & 969 & 7.80 & 970 & 7.71 & 969 & 7.80 \\
		 2650 & 2744 & 1147 & 1059 & 981 & 7.37 & 975 & 7.93 & 977 & 7.74 & 975 & 7.93 \\
		 2800 & 2853 & 1225 & 1137 & 1048 & 7.83 & 1048 & 7.83 & 1044 & 8.18 & 1040 & 8.53 \\
		%16 & 2850 & 2909 & 1218 & 1141 & 1057 & 7.36 & 1051 & 7.89 & 1054 & 7.62 & 1051 & 7.89 \\
		%17 & 2900 & 2986 & 1259 & 1163 & 1084 & 6.79 & 1074 & 7.65 & 1075 & 7.57 & 1074 & 7.65 \\
		 2950 & 3011 & 1277 & 1186 & 1091 & 8.01 & 1085 & 8.52 & 1089 & 8.18 & 1085 & 8.52 \\
		%19 & 3000 & 3077 & 1282 & 1187 & 1105 & 6.91 & 1098 & 7.50 & 1099 & 7.41 & 1098 & 7.50 \\
%		20 & 3250 & 3388 & 1405 & 1301 & 1215 & 6.61 & 1201 & 7.69 & 1208 & 7.15 & 1200 & 7.76 \\
		 3300 & 3342 & 1440 & 1339 & 1232 & 7.99 & 1219 & 8.96 & 1221 & 8.81 & 1219 & 8.96 \\
%		22 & 3350 & 3398 & 1446 & 1356 & 1267 & 6.56 & 1256 & 7.37 & 1259 & 7.15 & 1254 & 7.52 \\
		 3400 & 3518 & 1480 & 1382 & 1273 & 7.89 & 1268 & 8.25 & 1269 & 8.18 & 1268 & 8.25 \\
%		24 & 3450 & 3592 & 1503 & 1396 & 1296 & 7.16 & 1323 & 5.23 & 1282 & 8.17 & 1278 & 8.45 \\
		 3500 & 3530 & 1537 & 1423 & 1305 & 8.29 & 1292 & 9.21 & 1296 & 8.92 & 1292 & 9.21 \\
%		26 & 3600 & 3609 & 1583 & 1468 & 1355 & 7.70 & 1347 & 8.24 & 1349 & 8.11 & 1347 & 8.24 \\
		 3650 & 3705 & 1589 & 1464 & 1367 & 6.63 & 1351 & 7.72 & 1356 & 7.38 & 1351 & 7.72 \\
%		28 & 3900 & 4020 & 1683 & 1561 & 1450 & 7.11 & 1434 & 8.14 & 1437 & 7.94 & 1434 & 8.14 \\
		 3950 & 4093 & 1688 & 1575 & 1464 & 7.05 & 1453 & 7.75 & 1455 & 7.62 & 1453 & 7.75 \\
%		30 & 4050 & 4168 & 1746 & 1635 & 1524 & 6.79 & 1518 & 7.16 & 1517 & 7.22 & 1511 & 7.58 \\
%		31 & 4150 & 4160 & 1799 & 1683 & 1554 & 7.66 & 1545 & 8.20 & 1547 & 8.08 & 1545 & 8.20 \\
		 4200 & 4202 & 1832 & 1707 & 1584 & 7.21 & 1564 & 8.38 & 1565 & 8.32 & 1564 & 8.38 \\
%		33 & 4250 & 4299 & 1844 & 1715 & 1592 & 7.17 & 1578 & 7.99 & 1579 & 7.93 & 1578 & 7.99 \\
%		 4450 & 4477 & 1934 & 1795 & 1662 & 7.41 & 1649 & 8.13 & 1652 & 7.97 & 1649 & 8.13 \\
		 4550 & 4589 & 1963 & 1841 & 1701 & 7.60 & 1710 & 7.12 & 1701 & 7.60 & 1695 & 7.93 \\
%		36 & 4750 & 4813 & 2058 & 1927 & 1793 & 6.95 & 1785 & 7.37 & 1782 & 7.52 & 1778 & 7.73 \\
		 4850 & 4915 & 2111 & 1955 & 1809 & 7.47 & 1790 & 8.44 & 1796 & 8.13 & 1789 & 8.49 \\
		 4950 & 5048 & 2164 & 2005 & 1865 & 6.98 & 1852 & 7.63 & 1853 & 7.58 & 1849 & 7.78 \\
%		39 & 5000 & 5071 & 2168 & 2015 & 1872 & 7.10 & 1855 & 7.94 & 1858 & 7.79 & 1853 & 8.04 \\
		5150 & 5288 & 2204 & 2044 & 1899 & 7.09 & 1888 & 7.63 & 1890 & 7.53 & 1888 & 7.63 \\
%		41 & 5200 & 5251 & 2265 & 2100 & 1947 & 7.29 & 1948 & 7.24 & 1940 & 7.62 & 1935 & 7.86 \\
%		42 & 5250 & 5374 & 2239 & 2090 & 1940 & 7.18 & 1932 & 7.56 & 1937 & 7.32 & 1932 & 7.56 \\
		5300 & 5313 & 2293 & 2136 & 1999 & 6.41 & 1980 & 7.30 & 1982 & 7.21 & 1980 & 7.30 \\
%		44 & 5350 & 5386 & 2325 & 2168 & 2026 & 6.55 & 2005 & 7.52 & 2006 & 7.47 & 2005 & 7.52 \\
		 5450 & 5489 & 2369 & 2211 & 2067 & 6.51 & 2052 & 7.19 & 2055 & 7.06 & 2052 & 7.19 \\
%		46 & 5500 & 5531 & 2378 & 2222 & 2063 & 7.16 & 2055 & 7.52 & 2056 & 7.47 & 2050 & 7.74 \\
		 5550 & 5562 & 2424 & 2255 & 2084 & 7.58 & 2069 & 8.25 & 2074 & 8.03 & 2069 & 8.25 \\
%		48 & 5600 & 5610 & 2426 & 2245 & 2091 & 6.86 & 2073 & 7.66 & 2074 & 7.62 & 2073 & 7.66 \\
		 5700 & 5770 & 2463 & 2287 & 2130 & 6.86 & 2115 & 7.52 & 2119 & 7.35 & 2114 & 7.56 \\
%		50 & 5750 & 5781 & 2487 & 2331 & 2168 & 6.99 & 2165 & 7.12 & 2159 & 7.38 & 2151 & 7.72 \\
		5950 & 6047 & 2614 & 2439 & 2256 & 7.50 & 2243 & 8.04 & 2246 & 7.91 & 2243 & 8.04 \\
%		52 & 6000 & 6010 & 2614 & 2427 & 2241 & 7.66 & 2237 & 7.83 & 2232 & 8.03 & 2225 & 8.32 \\
		 6000 & 6010 & 2614 & 2427 & 2241 & 7.66 & 2237 & 7.83 & 2232 & 8.03 & 2225 & 8.32 \\
%		54 & 6100 & 6162 & 2642 & 2488 & 2298 & 7.64 & 2291 & 7.92 & 2297 & 7.68 & 2289 & 8.00 \\
		 6200 & 6310 & 2694 & 2490 & 2304 & 7.47 & 2309 & 7.27 & 2289 & 8.07 & 2282 & 8.35 \\
%		56 & 6350 & 6491 & 2722 & 2529 & 2340 & 7.47 & 2335 & 7.67 & 2331 & 7.83 & 2325 & 8.07 \\
		 6400 & 6466 & 2769 & 2582 & 2400 & 7.05 & 2395 & 7.24 & 2388 & 7.51 & 2383 & 7.71 \\
%		58 & 6450 & 6465 & 2841 & 2628 & 2432 & 7.46 & 2421 & 7.88 & 2429 & 7.57 & 2415 & 8.11 \\
		 6500 & 6586 & 2804 & 2623 & 2440 & 6.98 & 2421 & 7.70 & 2424 & 7.59 & 2421 & 7.70 \\
%		60 & 6650 & 6794 & 2875 & 2682 & 2494 & 7.01 & 2488 & 7.23 & 2483 & 7.42 & 2478 & 7.61 \\
		 6750 & 6870 & 2952 & 2721 & 2522 & 7.31 & 2494 & 8.34 & 2499 & 8.16 & 2494 & 8.34 \\
%		62 & 6800 & 6853 & 2945 & 2724 & 2539 & 6.79 & 2523 & 7.38 & 2524 & 7.34 & 2519 & 7.53 \\
		 6850 & 6933 & 2969 & 2785 & 2579 & 7.40 & 2566 & 7.86 & 2566 & 7.86 & 2563 & 7.97 \\
%		64 & 6900 & 6940 & 2988 & 2774 & 2576 & 7.14 & 2556 & 7.86 & 2557 & 7.82 & 2556 & 7.86 \\
		 7000 & 7038 & 3043 & 2832 & 2620 & 7.49 & 2623 & 7.38 & 2611 & 7.80 & 2601 & 8.16 \\
%		66 & 7050 & 7142 & 3055 & 2838 & 2646 & 6.77 & 2635 & 7.15 & 2627 & 7.43 & 2623 & 7.58 \\
		 7100 & 7135 & 3067 & 2851 & 2646 & 7.19 & 2639 & 7.44 & 2633 & 7.65 & 2621 & 8.07 \\
%		68 & 7200 & 7274 & 3110 & 2912 & 2691 & 7.59 & 2674 & 8.17 & 2678 & 8.04 & 2674 & 8.17 \\
		 7250 & 7357 & 3080 & 2879 & 2671 & 7.22 & 2654 & 7.82 & 2656 & 7.75 & 2654 & 7.82 \\
%		70 & 7300 & 7311 & 3159 & 2929 & 2722 & 7.07 & 2702 & 7.75 & 2706 & 7.61 & 2702 & 7.75 \\
		 7350 & 7474 & 3177 & 2944 & 2744 & 6.79 & 2717 & 7.71 & 2722 & 7.54 & 2717 & 7.71 \\
%		72 & 7400 & 7474 & 3200 & 2981 & 2768 & 7.15 & 2771 & 7.04 & 2742 & 8.02 & 2738 & 8.15 \\
		 7450 & 7497 & 3217 & 2980 & 2771 & 7.01 & 2772 & 6.98 & 2757 & 7.48 & 2751 & 7.68 \\
		 7550 & 7557 & 3299 & 3093 & 2844 & 8.05 & 2827 & 8.60 & 2835 & 8.34 & 2821 & 8.79 \\
		 7650 & 7696 & 3331 & 3095 & 2875 & 7.11 & 2849 & 7.95 & 2853 & 7.82 & 2846 & 8.05 \\
		 7700 & 7716 & 3332 & 3096 & 2875 & 7.14 & 2861 & 7.59 & 2862 & 7.56 & 2855 & 7.78 \\
		 7800 & 7804 & 3395 & 3176 & 2947 & 7.21 & 2933 & 7.65 & 2938 & 7.49 & 2924 & 7.93 \\
		 7950 & 7982 & 3480 & 3260 & 3022 & 7.30 & 3000 & 7.98 & 3004 & 7.85 & 3000 & 7.98 \\
		 8000 & 8126 & 3466 & 3215 & 2987 & 7.09 & 2985 & 7.15 & 2969 & 7.65 & 2957 & 8.02 \\
		 9900 & 24411032 & 16 & 14 & 13 & 7.14 & 13 & 7.14 & 13 & 7.14 & 13 & 7.14 \\
		 15000 & 56111357 & 14 & 12 & 12 & 0 & 12 & 0 & 12 & 0 & 12 & 0 \\ \hline
		\caption{Results for sample instances}
		\label{table1}
	\end{longtable}
	
	\normalsize

	Table \ref{table2} below presents the percentage of the improvement 
	accomplished by purification procedures 1-4, on average, for all the 
	tested benchmark instances.

	\begin{longtable}[c]{cccc}
		\hline
			$PP_1$	& $PP_2$	& $PP_3$ & $PP_4$\\
		\hline
			9.17		& 9.48			& 8.88 & 9.63\\
			
		\hline
		\caption{Average results over all instances \label{table2}}
	\end{longtable}
	\unskip

	In 	average over all the 1316 tested instances, a best solution among the 
	four solutions generated by purification procedures 1-4,  
	was 9.68\%  smaller than the dominant set returned by $PP_0$. % purification procedure 0. 
	Purification procedures 1-4 turned out to be more efficient for dance 
	instances with about 40\% of the improvement, though in absolute
	 terms, 	the number of purified vertices	for none-dance instances was higher.

\begin{figure}[H]
		\centering
		\includegraphics[width=0.65\linewidth]{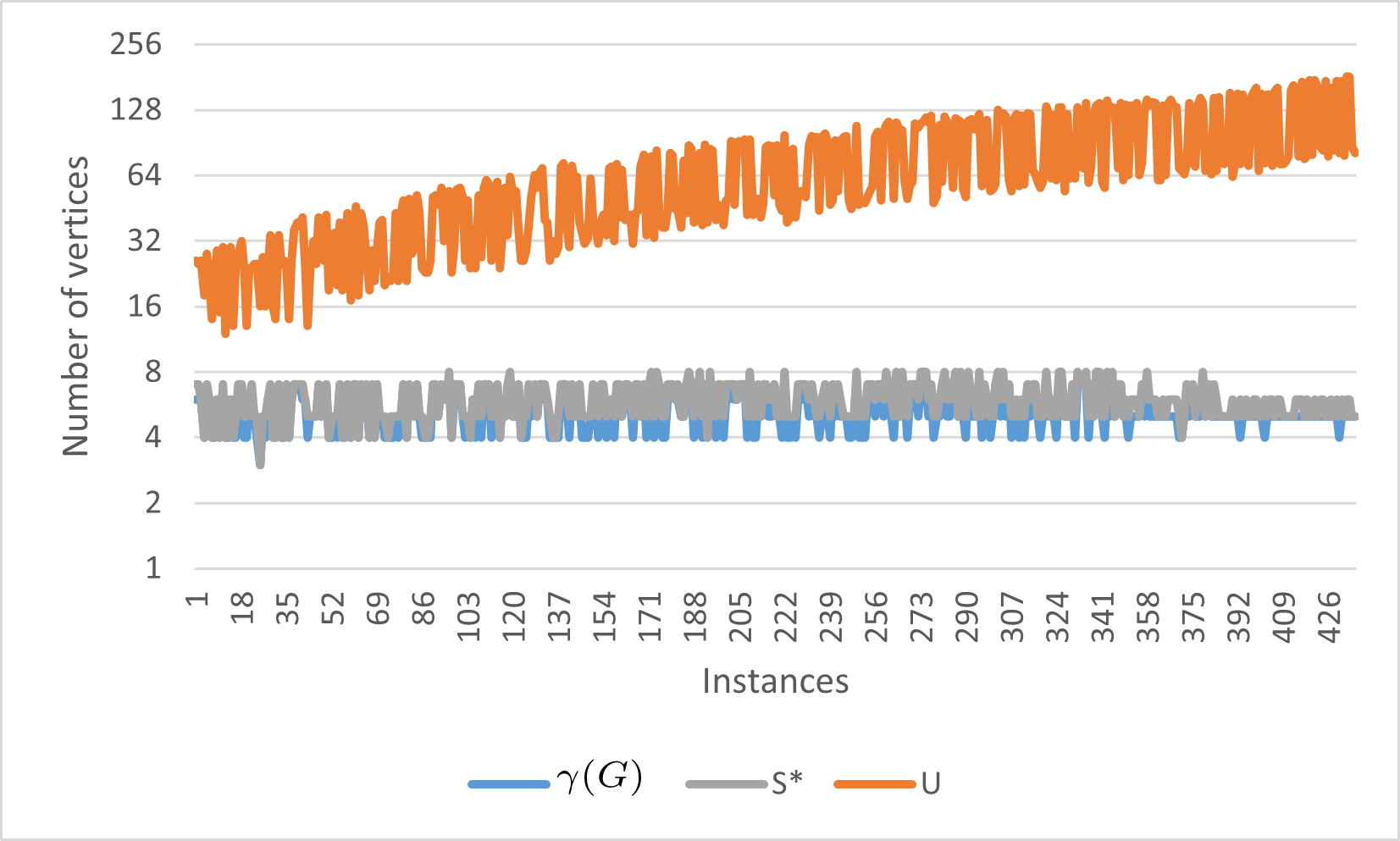}
		\caption{Results of purification procedures ($\min\{S^*_i\}$), $\gamma(G)$, and $ U $}
		\label{figura2}
	\end{figure}

	In Table \ref{table3} presents results for 45 sample benchmark instances from 
	\cite{bdparra2} for which the optima solutions are known. This table
	compares the quality of the solutions obtained by purification procedures 1-4 
	 vs the corresponding optimal solutions and upper bounds. The complete
	 comparative analysis for all the 500 benchmark instances with known optimal 
	  solutions can be found at \cite{bdparra2}, and is also reflected in
	  Figure \ref{figura2}.  Over all the tested instances, 
	 our solutions,  on average, contained 1/7th part of the number of vertices
	 established due  to the best known upper bound $U$, 	 an improvement of 
	 85.71\% (see Fact \ref{teorema1} at the end of this section). Remarkably, 
	 among the 500 instances 
	 with the known optimum, an optimal solution was generated for 
	 46.33\% of the instances, whereas the average error over all the 
	 created non-optimal solutions was 1.01 vertices.

	% Please add the following required packages to your document preamble:
	% \usepackage{multirow}
	% \usepackage[table,xcdraw]{xcolor}
	% If you use beamer only pass "xcolor=table" option, i.e. \documentclass[xcolor=table]{beamer}
	% \usepackage{longtable}
	% Note: It may be necessary to compile the document several times to get a multi-page table to line up properly
	\footnotesize
	%\scriptsize
	\begin{longtable}[c]{cccccccc}
		\hline
		  &  &  & $PP_1$ & $PP_2$ & $PP_3$ & $PP_4$ &  \\ \cline{4-7}
		 \multirow{-2}{*}{\textbf{$|V|$}} & \multirow{-2}{*}{\textbf{$|E|$}} & \multirow{-2}{*}{$\gamma(G)$} & \textbf{$|S^*_1|$} & \textbf{$|S^*_2|$} & \textbf{$|S^*_3|$} & \textbf{$|S^*_R|$} & \multirow{-2}{*}{\textbf{U}} \\ \hline
		\endhead
		\hline
		\endfoot
		\endlastfoot
100 & 2687 & 6 & 7 & 7 & 7 & 7 & 26 \\
106 & 3744 & 4 & 4 & 4 & 4 & 4 & 18 \\
126 & 4651 & 6 & 6 & 6 & 6 & 6 & 30 \\
128 & 6331 & 4 & 4 & 4 & 4 & 4 & 13 \\
134 & 5359 & 7 & 7 & 7 & 7 & 7 & 32 \\
138 & 7470 & 4 & 5 & 5 & 5 & 5 & 13 \\
140 & 6710 & 6 & 6 & 6 & 6 & 6 & 24 \\
144 & 7267 & 5 & 5 & 5 & 5 & 5 & 25 \\
148 & 8664 & 3 & 3 & 3 & 4 & 3 & 16 \\
150 & 7950 & 5 & 5 & 5 & 5 & 5 & 27 \\
154 & 8304 & 6 & 6 & 6 & 6 & 6 & 24 \\
170 & 11640 & 4 & 4 & 4 & 4 & 4 & 14 \\
172 & 10665 & 6 & 6 & 6 & 6 & 6 & 26 \\
176 & 9978 & 7 & 7 & 7 & 7 & 7 & 39 \\
182 & 12100 & 6 & 6 & 6 & 6 & 6 & 25 \\
194 & 13675 & 5 & 5 & 5 & 5 & 5 & 34 \\
196 & 14229 & 6 & 6 & 6 & 6 & 6 & 26 \\
198 & 12936 & 7 & 7 & 7 & 7 & 7 & 42 \\
200 & 16389 & 4 & 4 & 4 & 4 & 4 & 19 \\
204 & 15169 & 5 & 5 & 5 & 5 & 5 & 35 \\
222 & 20370 & 4 & 4 & 4 & 4 & 4 & 18 \\
224 & 16923 & 7 & 7 & 7 & 7 & 7 & 43 \\
234 & 22709 & 4 & 4 & 4 & 4 & 4 & 21 \\
238 & 20902 & 5 & 5 & 5 & 5 & 5 & 39 \\
264 & 27175 & 5 & 5 & 5 & 6 & 5 & 30 \\
326 & 37878 & 7 & 7 & 7 & 7 & 7 & 60 \\
328 & 45703 & 4 & 4 & 4 & 4 & 4 & 24 \\
330 & 43714 & 6 & 6 & 6 & 6 & 6 & 35 \\
332 & 41661 & 5 & 5 & 5 & 5 & 5 & 56 \\
342 & 47142 & 7 & 7 & 7 & 7 & 7 & 36 \\
344 & 50364 & 4 & 4 & 4 & 4 & 4 & 26 \\
370 & 58531 & 4 & 4 & 4 & 4 & 4 & 28 \\
400 & 65545 & 6 & 6 & 6 & 6 & 6 & 37 \\
458 & 77296 & 7 & 7 & 7 & 7 & 7 & 79 \\
506 & 95091 & 6 & 6 & 6 & 7 & 6 & 91 \\
538 & 125734 & 5 & 5 & 5 & 5 & 5 & 47 \\
612 & 140918 & 7 & 7 & 7 & 8 & 7 & 101 \\
714 & 198539 & 5 & 5 & 5 & 5 & 5 & 117 \\
724 & 199141 & 7 & 7 & 7 & 7 & 7 & 124 \\
790 & 243725 & 5 & 5 & 5 & 5 & 5 & 130 \\
794 & 240592 & 7 & 7 & 7 & 7 & 7 & 138 \\
840 & 310396 & 4 & 4 & 4 & 5 & 4 & 67 \\
844 & 306681 & 7 & 7 & 7 & 7 & 7 & 72 \\
870 & 296197 & 5 & 7 & 7 & 7 & 7 & 145 \\
1098 & 533220 & 5 & 5 & 5 & 5 & 5 & 81 \\ \hline
		\caption{Comparison between results of purification procedures, $\gamma(G)$, and $ U $ \label{table3}}
		
	\end{longtable}
	\normalsize
	
\begin{figure}[H]
		\centering
		\begin{subfigure}[b]{0.45\textwidth}
			\centering
			\includegraphics[width=\textwidth]{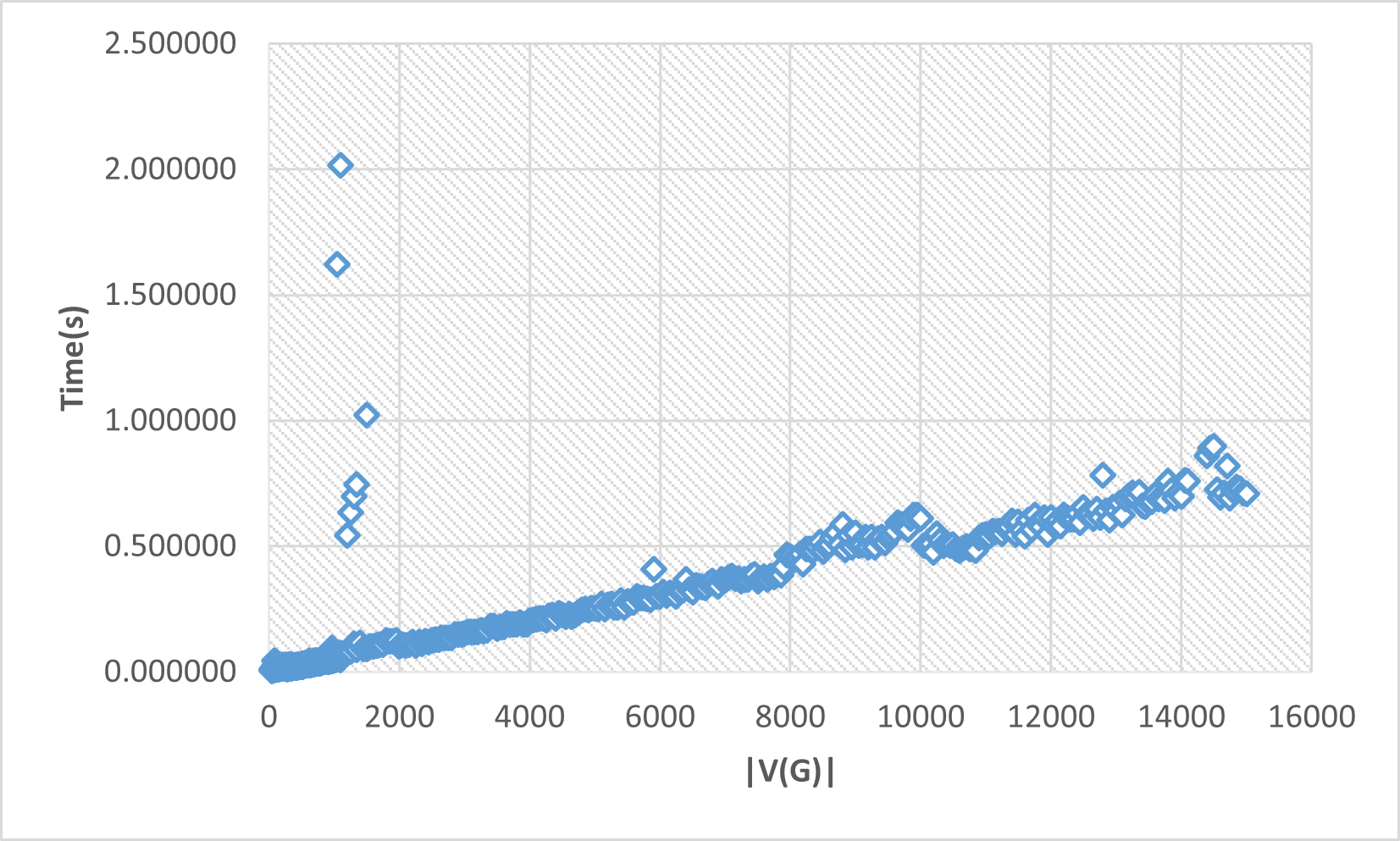}
			\caption{Maximum execution times among $ PP_{1-4} $ for all instances}
			\label{figura3a}
		\end{subfigure}
		\hfill
		\begin{subfigure}[b]{0.45\textwidth}
			\centering
			\includegraphics[width=\textwidth]{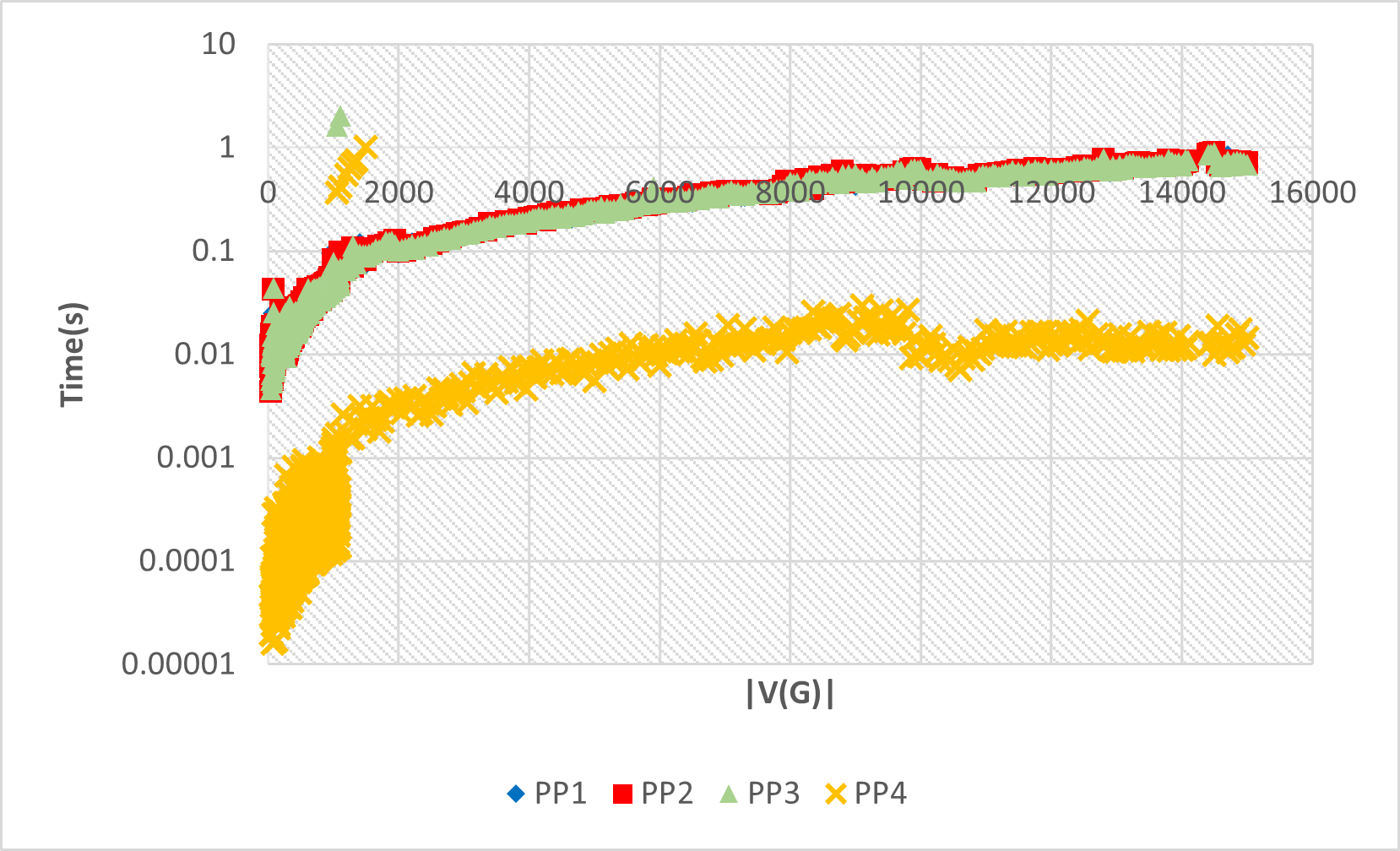}
			\caption{Execution times for individual procedures}
			\label{figura3b}
		\end{subfigure}
		\caption{Time of purification procedures}
		\label{figura3}
	\end{figure}

	As to  the execution times, for the largest benchmark instances with
	15000 vertices, all four procedures finished in less than two seconds.   
	Figure \ref{figura3} represents the execution times for all the tested
	instances.

\begin{figure}[H]
	\centering
	\includegraphics[width=1.0\linewidth]{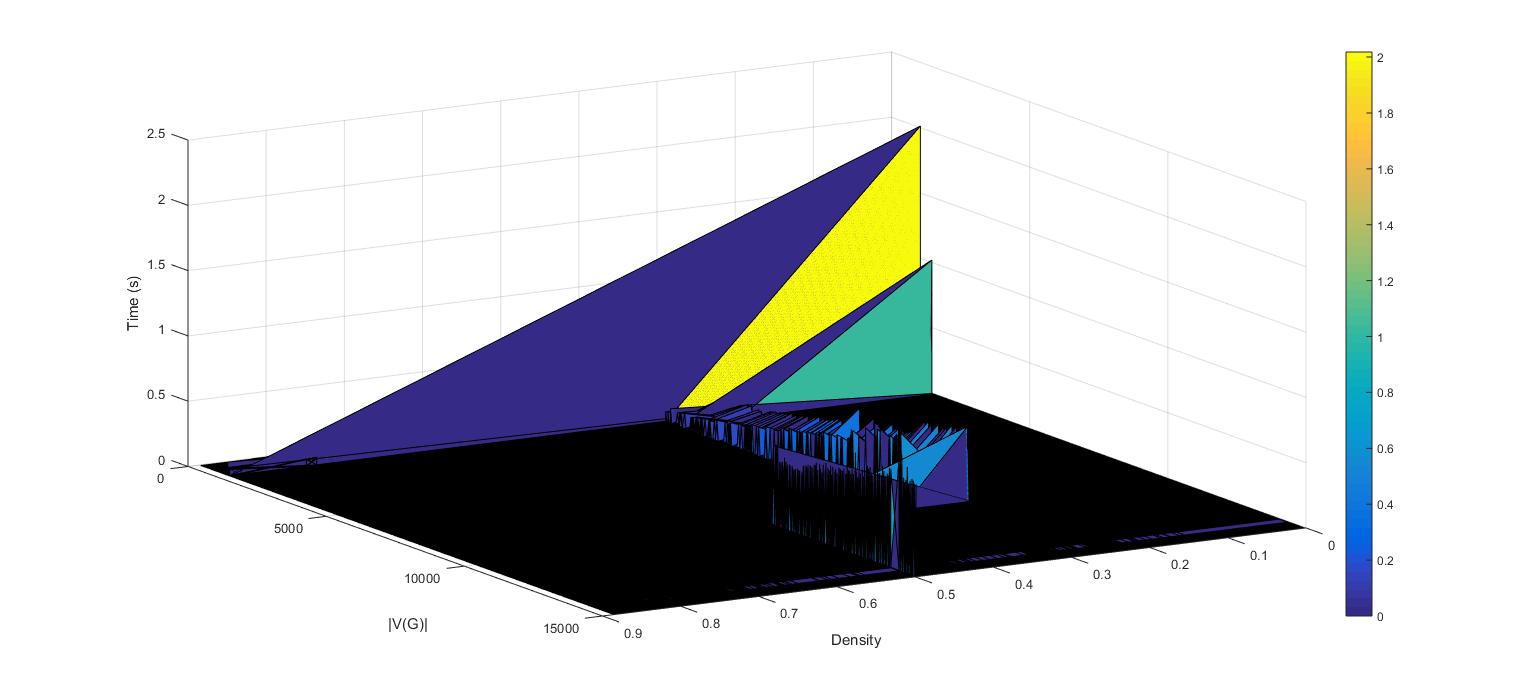}
	\caption{Problem instance density and size vs the corresponding 
	maximum execution time}
	\label{figura4}
\end{figure}

 Figure \ref{figura4} below reflects the dependence of execution times
 on the density and the size of the instances.

	The upper bound $U$ is due to the following known results. 
	
	\begin{fact}\label{teorema1}
		\cite{haynes1998domination} If $G(V,E)$ is a connected graph of order $n$, minimum degree $\delta(G)$ and maximum degree $\Delta(G)$, then $ \gamma(G)\leq \frac{n}{2} $, $ \gamma(G)\leq n-\Delta(G)$, and $ \gamma(G)\leq \frac{n(\ln(\delta(G)+1))}{\delta(G)+1} $. Hence, $$U=\min\left\lbrace \frac{n}{2}, n-\Delta(G), \frac{n(\ln(\delta(G)+1))}{\delta(G)+1}\right\rbrace $$ 
		is a valid upper bound on $\gamma(G)$.		
 
	\end{fact}

%	 \section{Acknowledgments} 
%	 
%	 The purification procedure $PP_4$ described in this note was originally 
%	 proposed by one of the anonymous referees of the reference \cite{tcs22}. 
%	 The authors express their gratitude to this referee. 

		%=====================================
		% References, variant A: external bibliography
		%=====================================
		\bibliography{mybibfile_purification}

\begin{thebibliography}{10}
\expandafter\ifx\csname url\endcsname\relax
  \def\url#1{\texttt{#1}}\fi
\expandafter\ifx\csname urlprefix\endcsname\relax\def\urlprefix{URL }\fi
\expandafter\ifx\csname href\endcsname\relax
  \def\href#1#2{#2} \def\path#1{#1}\fi

\bibitem{Garey}
M.~R. Garey, D.~S. Johnson, Computers and intractability, Vol. 174, freeman San
  Francisco, 1979.

\bibitem{berge1962theory}
C.~Berge, The theory of graphs and its applications, Methuen \& Co, Ltd.,
  London.

\bibitem{ore1962theory}
O.~Ore, Theory of graphs, ams colloquium publications 38, Publ. Providence,
  Rhode Island.

\bibitem{haynes1998domination}
T.~W. Haynes, S.~T. Hedetniemi, P.~J. Slater, Domination in graphs (advanced
  topics) marcel dekker publications, New York.

\bibitem{haynes2017domination}
T.~W. Haynes, Domination in Graphs: Volume 2: Advanced Topics, Routledge, 2017.

\bibitem{CORCORAN2021105368}
P.~Corcoran, A.~Gagarin, Heuristics for k-domination models of facility
  location problems in street networks, Computers \& Operations Research 133
  (2021) 105368.
\newblock \href {http://dx.doi.org/https://doi.org/10.1016/j.cor.2021.105368}
  {\path{doi:https://doi.org/10.1016/j.cor.2021.105368}}.

\bibitem{JOSHI1994352}
D.~S. Joshi, S.~Radhakrishnan, N.~Chandrasekharan, The k-neighbor, r-domination
  problems on interval graphs, European Journal of Operational Research 79~(2)
  (1994) 352--368.
\newblock \href
  {http://dx.doi.org/https://doi.org/10.1016/0377-2217(94)90364-6}
  {\path{doi:https://doi.org/10.1016/0377-2217(94)90364-6}}.

\bibitem{liao2005power}
C.-S. Liao, D.-T. Lee, Power domination problem in graphs, in: International
  Computing and Combinatorics Conference, Springer, 2005, pp. 818--828.
\newblock \href {http://dx.doi.org/https://doi.org/10.1007/11533719_83}
  {\path{doi:https://doi.org/10.1007/11533719_83}}.

\bibitem{haynes2002domination}
T.~W. Haynes, S.~M. Hedetniemi, S.~T. Hedetniemi, M.~A. Henning, Domination in
  graphs applied to electric power networks, SIAM journal on discrete
  mathematics 15~(4) (2002) 519--529.
\newblock \href {http://dx.doi.org/https://doi.org/10.1137/S0895480100375831}
  {\path{doi:https://doi.org/10.1137/S0895480100375831}}.

\bibitem{RTParra}
E.~Parra~Inza, A.~Sandoval~Ramírez, J.~Hernández~Gómez, G.~Cerdenares
  Ladrón~de Guevara, Robustness Analysis of Trophic Networks using Outer
  k-independent Total Dominant Sets. In Modelación Matemática IV
  Biomatemáticas, Epidemiología, Ingeniería., Vol.~4, Universidad
  Tecnológica de la Mixteca, 2021.

\bibitem{balasundaram2006graph}
B.~Balasundaram, S.~Butenko, Graph domination, coloring and cliques in
  telecommunications, in: Handbook of optimization in telecommunications,
  Springer, 2006, pp. 865--890.
\newblock \href
  {http://dx.doi.org/https://doi.org/10.1007/978-0-387-30165-5_30}
  {\path{doi:https://doi.org/10.1007/978-0-387-30165-5_30}}.

\bibitem{wan2004distributed}
P.-J. Wan, K.~M. Alzoubi, O.~Frieder, Distributed construction of connected
  dominating set in wireless ad hoc networks, Mobile Networks and Applications
  9~(2) (2004) 141--149.
\newblock \href
  {http://dx.doi.org/https://doi.org/10.1023/B:MONE.0000013625.87793.13}
  {\path{doi:https://doi.org/10.1023/B:MONE.0000013625.87793.13}}.

\bibitem{wu2003power}
J.~Wu, B.~Wu, I.~Stojmenovic, Power-aware broadcasting and activity scheduling
  in ad hoc wireless networks using connected dominating sets, Wireless
  Communications and Mobile Computing 3~(4) (2003) 425--438.
\newblock \href {http://dx.doi.org/https://doi.org/10.1002/wcm.125}
  {\path{doi:https://doi.org/10.1002/wcm.125}}.

\bibitem{wu2002extended}
J.~Wu, Extended dominating-set-based routing in ad hoc wireless networks with
  unidirectional links, IEEE transactions on parallel and distributed systems
  13~(9) (2002) 866--881.
\newblock \href {http://dx.doi.org/https://doi.org/10.1109/TPDS.2002.1036062}
  {\path{doi:https://doi.org/10.1109/TPDS.2002.1036062}}.

\bibitem{wang2009}
F.~Wang, E.~Camacho, K.~Xu, Positive influence dominating set in online social
  networks, in: Combinatorial Optimization and Applications: Third
  International Conference, COCOA 2009, Huangshan, China, June 10-12, 2009.
  Proceedings 3, Springer, 2009, pp. 313--321.
\newblock \href
  {http://dx.doi.org/https://doi.org/10.1007/978-3-642-02026-1_29}
  {\path{doi:https://doi.org/10.1007/978-3-642-02026-1_29}}.

\bibitem{wang2011}
F.~Wang, H.~Du, E.~Camacho, K.~Xu, W.~Lee, Y.~Shi, S.~Shan, On positive
  influence dominating sets in social networks, Theoretical Computer Science
  412~(3) (2011) 265--269.
\newblock \href {http://dx.doi.org/https://doi.org/10.1016/j.tcs.2009.10.001}
  {\path{doi:https://doi.org/10.1016/j.tcs.2009.10.001}}.

\bibitem{abu2018}
F.~N. Abu-Khzam, K.~Lamaa, Efficient heuristic algorithms for
  positive-influence dominating set in social networks, in: IEEE INFOCOM
  2018-IEEE Conference on Computer Communications Workshops (INFOCOM WKSHPS),
  IEEE, 2018, pp. 610--615.
\newblock \href
  {http://dx.doi.org/https://doi.org/10.1109/INFCOMW.2018.8406851}
  {\path{doi:https://doi.org/10.1109/INFCOMW.2018.8406851}}.

\bibitem{chvatal}
V.~Chvatal, A greedy heuristic for the set-covering problem, Mathematics of
  operations research 4~(3) (1979) 233--235.
\newblock \href {http://dx.doi.org/https://doi.org/10.1287/moor.4.3.233}
  {\path{doi:https://doi.org/10.1287/moor.4.3.233}}.

\bibitem{parekh}
A.~K. Parekh, Analysis of a greedy heuristic for finding small dominating sets
  in graphs, Information processing letters 39~(5) (1991) 237--240.
\newblock \href
  {http://dx.doi.org/https://doi.org/10.1016/0020-0190(91)90021-9}
  {\path{doi:https://doi.org/10.1016/0020-0190(91)90021-9}}.

\bibitem{eubank}
S.~Eubank, V.~A. Kumar, M.~V. Marathe, A.~Srinivasan, N.~Wang, Structural and
  algorithmic aspects of massive social networks, in: Proceedings of the
  fifteenth annual ACM-SIAM symposium on Discrete algorithms, 2004, pp.
  718--727.

\bibitem{campan}
A.~Campan, T.~M. Truta, M.~Beckerich, Fast dominating set algorithms for social
  networks., in: MAICS, 2015, pp. 55--62.

\bibitem{tcs22}
F.~{Hernández Mira}, E.~{Parra Inza}, J.~M. {Sigarreta Almira}, N.~Vakhania, A
  polynomial-time approximation to a minimum dominating set in a graph,
  Theoretical Computer Science 930 (2022) 142--156.
\newblock \href {http://dx.doi.org/https://doi.org/10.1016/j.tcs.2022.07.020}
  {\path{doi:https://doi.org/10.1016/j.tcs.2022.07.020}}.

\bibitem{van2011exact}
J.~M. Van~Rooij, H.~L. Bodlaender, Exact algorithms for dominating set,
  Discrete Applied Mathematics 159~(17) (2011) 2147--2164.
\newblock \href {http://dx.doi.org/https://doi.org/10.1016/j.dam.2011.07.001}
  {\path{doi:https://doi.org/10.1016/j.dam.2011.07.001}}.

\bibitem{iwata2012faster}
Y.~Iwata, A faster algorithm for dominating set analyzed by the potential
  method, in: International Symposium on Parameterized and Exact Computation,
  Springer, 2012, pp. 41--54.
\newblock \href {http://dx.doi.org/https://doi.org/10.1007/978-3-642-28050-4_4}
  {\path{doi:https://doi.org/10.1007/978-3-642-28050-4_4}}.

\bibitem{Parra2022}
E.~{Parra Inza}, N.~Vakhania, J.~M. {Sigarreta Almira}, F.~A. {Hernández
  Mira}, Exact and heuristic algorithms for the domination problem, European
  Journal of Operational Research 313~(3) (2024) 926--936.
\newblock \href {http://dx.doi.org/https://doi.org/10.1016/j.ejor.2023.08.033}
  {\path{doi:https://doi.org/10.1016/j.ejor.2023.08.033}}.

\bibitem{bdparra2}
E.~Parra~Inza, Random graph (1), Mendeley Data V4.
\newblock \href {http://dx.doi.org/https://doi.org/10.17632/rr5bkj6dw5.4}
  {\path{doi:https://doi.org/10.17632/rr5bkj6dw5.4}}.

\end{thebibliography}

\end{document}